\documentclass[twocolumn,showpacs,preprintnumbers,amsmath,amssymb]{revtex4}

\usepackage{graphicx}
\usepackage{dcolumn}
\usepackage{bm}
\usepackage{color}
\def\be{\begin{equation}}
\def\ee{\end{equation}}
\def\bea{\begin{eqnarray}}
\def\eea{\end{eqnarray}}
\def\la{\langle}
\def\ra{\rangle}

\begin{document}


\title{Heat Conduction Process on Community Networks as a Recommendation Model}

\author{Yi-Cheng Zhang\footnotemark{} and Marcel Blattner}
 \affiliation{Physics Department, University of Fribourg, 1700 Fribourg, Switzerland \\
and Physics Department, Renmin University, Beijing, China}
\author{Yi-Kuo Yu\footnote{Corresponding author, email:  
 yyu@ncbi.nlm.nih.gov\\
\phantom{1}\hspace{79pt} $^*$email: yi-cheng.zhang@unifr.ch}}
\affiliation{National Center for Biotechnology Information, National 
 Library of Medicine, National Institutes of Health, Bethesda, MD 20894, USA}
   
\date{May 3rd, 2007}

\begin{abstract}
Using heat conduction mechanism on a social network we develop a systematic method 
to predict missing values as recommendations. This method can treat very large
  matrices that are typical of internet communities. In particular,
 with an innovative, exact formulation that accommodates arbitrary boundary condition,  
 our method is easy to use in real applications. The performance is assessed by 
comparing with traditional recommendation methods using real data.
\end{abstract}

\pacs{44.10.+i, 89.70.+c, 89.20.Hh}

\maketitle

With the advent of the internet, there sprout many web sites that
enable large communities to aggregate and interact.  For example
livejournal.com allows its 3 million members to share interests
and life experiences; del.icio.us is a social bookmark service for
 people to share their findings on the World Wide Web.
 Thousands of such web sites are built by web entrepreneurs
and activists for the public, and their number is growing ever
faster. This brings about massive amount of accessible
information, more than each individual is able or willing to process.
Information search, filtering, and recommendation thus 
 become indispensable in internet era. Ideally speaking, 
a good recommendation mechanism should be able to ``guess'' what a
person may want to select based on what he or she already
selected \cite{ZhMa01,BlZhMa06}. Many such mechanisms are in actual use (like
www.amazon.com proposing its readers with new books), however, jury
is still out as to what is the best model. 
For a review of current
techniques, see~\cite{AdTu05}. 

Based on the heat conduction (or diffusion) process, 
 we propose a recommendation model 
 capable of handling individualized boundary conditions (BC). 
 To better explain our model,  we first illustrate using the friendship 
 network of $N$ people:  
 each person (member) is a node, and a pair of nodes is connected 
 by an edge provided they are mutual friends. The collection
 of these information forms the {\it symmetric} adjacency matrix $A$: element  
 $A_{ij}=1$, or $0$ depending on whether people $i$ and $j$ 
 are mutual friends (1) or not (0). Although it is possible to consider 
 asymmetric connection, this generalization will not be studied here.
 To recommend friends to any individual member, we first set (Dirichlet)
BC: to set the values on the directly connected
nodes as 1 and some remote nodes (will be further specified)
as 0. Values on all other nodes are treated as variables to be
determined. These 
values can be interpreted as the probabilities that
these nodes might be selected as friends.

We now describe an efficient and effective strategy to solve the proposed
  heat conduction problem. From $A$, we first construct a  propagator matrix 
 $P= D^{-1} A $, where $D$ is the diagonal degree matrix. Denote $H$ as the temperature 
vector of $N$ components: the source-components are {\it high temperature}
 nodes with temperature $1$;
the sink-components are low temperature nodes with temperature $0$. 
  Our task is to find, through thermal equilibrium, the temperatures associated 
 with the remaining nodes that are neither sinks nor sources.  
 The discrete Laplace operator, analog of $-\nabla^2$, on this network is
 $L=I-P$, where $I$ is the identity matrix. We only need to solve  
\begin{equation} 
LH=f \label{eq:heat}
\end{equation}
where $f$ is the external {\it flux} vector. Note that this is 
 the discrete analog of
 $- \kappa \nabla^2 T(\vec r)  =  \nabla \cdot {\vec J}(\vec r)$
 with $H(i)$ plays the role of $\kappa T(\vec r)$ and 
 $f(i)$ plays the role of $\nabla \cdot {\vec J}(\vec r)$.

  Because Laplace operator conserves total heat and tend to spread
 heat from high temperature region to low temperature region, 
 the only way to maintain the fixed temperature values at
the sources and sinks is to apply external heat flux (inflow at
sources and outflow at sinks). For the rest of the nodes, 
 the equilibrium condition demands that no net heat flux should occur.
 Therefore, the only allowed nonzero components of $f$ 
 are source- and sink-components.  

The computation of the temperature vector is straightforward. It is
convenient to group the source and sink components together
into a block $H_{1}$, and the rest free variables another block
$H_{2}$. That is 
\be 
H = \left( \begin{array}{c}
H_1 \\
H_2 \end{array} \right) \; . \label{eq:H}
\ee

Likewise, we group the Laplace operator in a similar fashion and
 eq.~(\ref{eq:heat}) may be expressed as 
\begin{equation} 
\left ( \begin{array}{ll}
  L_{11} & L_{12}  \\ L_{21} & L_{22} \end{array} \right) 
 \left( \begin{array}{c} 
 H_1 \\ H_2 \end{array} \right) = 
 \left( \begin{array}{c} 
 f \\ 0 \end{array} \right) 
 \; .
\label{eq:L} 
\end{equation} 
All we need to solve is the homogeneous
equation 
\begin{equation} 
L_{21}H_{1} + L_{22}H_{2}=0 \; ,
\end{equation}
without the need to know $f$. Fixing the values of $H_1$, 
 $H_2$ can be readily found
using standard iterative methods~\cite{Num92}.
The above approach, although straightforward, 
represents a daunting challenge: for each individual, 
 we must solve the huge matrix problem once -- a 
 prohibitively expensive task for a typical internet
community having millions of members.

The standard way to get around this dilemma is to resort to
the Green's function method. Starting from eq.(\ref{eq:heat})
we would like to have a Green's function $\Omega'$ such that
eq.(\ref{eq:heat}) can be inverted: 
\begin{equation} 
\left( \begin{array}{c}
 H_1 \\ H_2 \end{array} \right) = \Omega' 
\left( \begin{array}{c} f \\ 0\end{array} \right) \; 
\end{equation} 
 to get $H_2 = \Omega'_{21}{\Omega'_{11}}^{-1} H_1$. 
However, $\Omega'=L^{-1} = (I-P)^{-1}$ is divergent:
 the Laplace operator has a zero eigenvalue and the inverse $L^{-1}$ is
 meaningful only if $(H_1,  H_2)^T$ is in the subspace that is orthogonal
 to the eigenvector of zero eigenvalue. A fortunate scenario like this has
 occurred in the studies of random resistor networks~\cite{Korniss_06,Wu_04}.

To simultaneously deal with all possible BC, we lose  
  the freedom to limit the solution to a certain subspace.
Nevertheless, we have a good understanding regarding this divergence.
 Basically, the $P$ matrix has an eigenvalue one with the right eigenvector
 being a column of $1$s
\[
|u^0\rangle = \left( 1, 1,  \cdots, 1 
\right)^{T}
\]
and with left eigenvector being 
\[
\langle v^0 | = \left( {d_1\over d}, {d_2\over d}, \ldots, {d_N \over d}
\right) 
\]
where $d_i$ denotes the degree of node $i$ and $d = \sum_i d_i$ being 
 the sum of degrees. Note that with this notation, we 
 have $\langle v^0 | u^0 \rangle = 1$.

We may then decompose $P$ into 
\[
P = Q + |u^0 \rangle \langle v^0 |
\]
with $Q |u^0 \rangle \langle v^0 | = |u^0 \rangle \langle v^0 | Q = 0$.
 Further, the spectral radius of $Q$ is now guaranteed to be smaller than
 $1$ and thus $(I-Q)$ is invertible with $(I-Q)^{-1} = \sum_{n=0}^\infty
 Q^n$. 
We may then rewrite the eq.(\ref{eq:L}) as 
\bea 
(I-Q)
\left(\begin{array}{c}
H_1 \\ H_2 \end{array} \right) &=& \left( \begin{array}{c} 
 f \\ 0 \end{array} \right)
+ |u^0 \rangle \langle v^0 |\left(\begin{array}{c}
H_1 \\ H_2 \end{array} \right) \nonumber \\
&=& \left( \begin{array}{c}  f \\ 0 \end{array} \right)
+ c(H) |u^0 \rangle \label{eq:L_1} 
\eea
 where the $H$-dependent constant may be written as 
$c(H) = \la v_1^0 | H_1 \ra + \la v_2^0 | H_2 \ra$.
 We need to  explain the notation further.  
 Basically $|u^0_1\ra$, represents a column vector
 whose components are obtained from the column vector $|u^0\ra$ 
 with component labels corresponding to that of the sources
 and the sinks. On the other hand, $|u^0_2\ra$ represents a column
 vector that is the remainder of $|u^0\ra$ after removing the
 components whose labels correspond to the sources and sinks. 
 Similarly, we define $\la v^0_1 | $ to be a row vector   
 whose components are obtained from the row vector $\la v^0 |$ 
 with component labels corresponding to that of the sources
 and the sinks; while $\la v^0_2 |$ represents a row
 vector that is the remainder of $ \la v^0 | $ after removing the
 components whose labels correspond to the sources and sinks. 
To simplify the notation, we will represent 
$c(H)$ by $c$ without explicitly showing its $H$ dependence. 

 Note that since $Q|u^0\ra = 0$, upon multiplying $\Omega \equiv(I-Q)^{-1}$
  to both side of eq.(\ref{eq:L_1}) we have
\be \label{sol.1}
\left(\begin{array}{c} 
H_1 \\ H_2 \end{array} \right) 
 = \left( \begin{array}{cc}
 \Omega_{11} & \Omega_{12} \\
\Omega_{21} & \Omega_{22} \end{array} \right)
 \left(\begin{array}{c} 
f \\ 0 \end{array} \right) + c|u_0\rangle 
\ee  
or equivalently
\be  \label{sol.2}
\left(\begin{array}{c} 
H_1-cu^0_1 \\ H_2-cu^0_2 \end{array} \right) 
 = \left( \begin{array}{cc}
 \Omega_{11} & \Omega_{12} \\
\Omega_{21} & \Omega_{22} \end{array} \right)
 \left(\begin{array}{c} 
f \\ 0 \end{array} \right)\; .
\ee
Consequently, we may write $H_2$ in the 
 following form
\be \label{sol.3}
| H_2 \ra = c \, |u^0_2 \ra + \Omega_{21} \Omega_{11}^{-1} |H_1 \ra 
 - c\,  \Omega_{21} \Omega_{11}^{-1} |u^0_1 \ra \; .
\ee 
Using the definition that $c = \la v^0_1 | H_1 \ra + \la v^0_2 | H_2 \ra $,
 we obtain
\[
c = \la v^0_1 | H_1\ra +  \la v^0_2 | \Omega_{21} \Omega_{11}^{-1} | H_1 \ra
+  c \left[ \la v_2^0 | u^0_2 \ra - \la v^0_2 | 
 \Omega_{21} \Omega_{11}^{-1} | u^0_1 \ra \right], 
\] 
or equivalently
\be \label{c.sol}
c = {  \la v^0_1 | H_1\ra +  \la v^0_2 | \Omega_{21} \Omega_{11}^{-1} | H_1 \ra
 \over 1 - \left[ \la v_2^0 | u^0_2 \ra - \la v^0_2 | 
 \Omega_{21} \Omega_{11}^{-1} | u^0_1 \ra \right] }
\ee
Substituting this result back to eq.~(\ref{sol.3}), we obtain 
  $H_2$ with computational complexity solely depending on 
 $\Omega_{21}\Omega_{11}^{-1}$. 
Note that we only needs to invert the matrix $(I-Q)$ once and for all. 
Upon specifying the boundary nodes, one needs to reshuffle 
the rows and columns of the matrix as well as vectors -- a relatively 
 efficient operation. This 
 operation groups the source nodes and sink nodes in one block
  to make easy the computation of $\Omega_{11}^{-1}$.  

 Let us emphasize that our final expression is written in a 
 rather general setting that it can be applied to cases
 when $P$ is either row-normalized or column-normalized.
In the case of column-normalized $P$, we will have  
 $|u_{\rm col.~norm.}^0\ra = (\la v^0_{\rm row~norm.} | )^T$
 and $\la v_{\rm col.~norm.}^0 | = (|u^0_{\rm row~norm.} \ra )^T$. 
The solution structures (\ref{sol.3}-\ref{c.sol}), however,
 does not change.

Although an exact Green's function method with Dirichlet boundary 
 condition using spectral analysis (eigenvalues and eigenvectors)
 has been established by Chung and Yau~\cite{ChYo00}, we find our 
 method more convenient for computational purpose. 	
With our method, the Greens function $\Omega$ is 
 computed once and can be used for all different BC. 
  This is immensely more efficient than finding all the
eigenvalues and eigenvectors for {\it every} BC needed
 for {\it each} individual. Furthermore, it would not be practical 
to find all the eigenvectors of matrices resulting from networks of 
 millions of nodes. 

To apply our method, one may either choose to fully invert $(I-Q)$ 
 or take its approximate form.
 The direct inversion of $(I-Q)$ may still be computationally challenging for
  a matrix of size millions by millions. In terms of approximations, 
we find the use of $(I-Q)^{-1} \equiv \lim_{M\to \infty} \Omega(M)$
particularly useful, with   
\be
 \Omega(M) \equiv \left[ I+  P+ \cdots + P^M - M|u_0\rangle \langle v_0 |
 \right] \; .
\ee 
This approximation gets better for larger $M$. This is because 
 the larger $M$ is, the smaller the difference between $P^M$
 and $|u_0\rangle \langle v_0 |$.  One may then 
 use $\Omega_{21}(M) {\Omega_{11}}^{\!\!\!\! -1}(M)$
 in place of $\Omega_{21} {\Omega_{11}}^{\!\!\!\! -1}$.  
 The quality of this approximation may be
 be verified comparing
 the two models: the exact solution(\ref{sol.3}-\ref{c.sol}) versus the
 approximate one (ie. replacing $\Omega_{21} {\Omega_{11}}^{\!\!\!\! -1}$
 by $\Omega_{21}(M) {\Omega_{11}}^{\!\!\!\! -1}(M)$ in the exact solution).

\begin{figure}
\vspace{-0.25in} 
\begin{center} 
\includegraphics[width=0.5\textwidth]{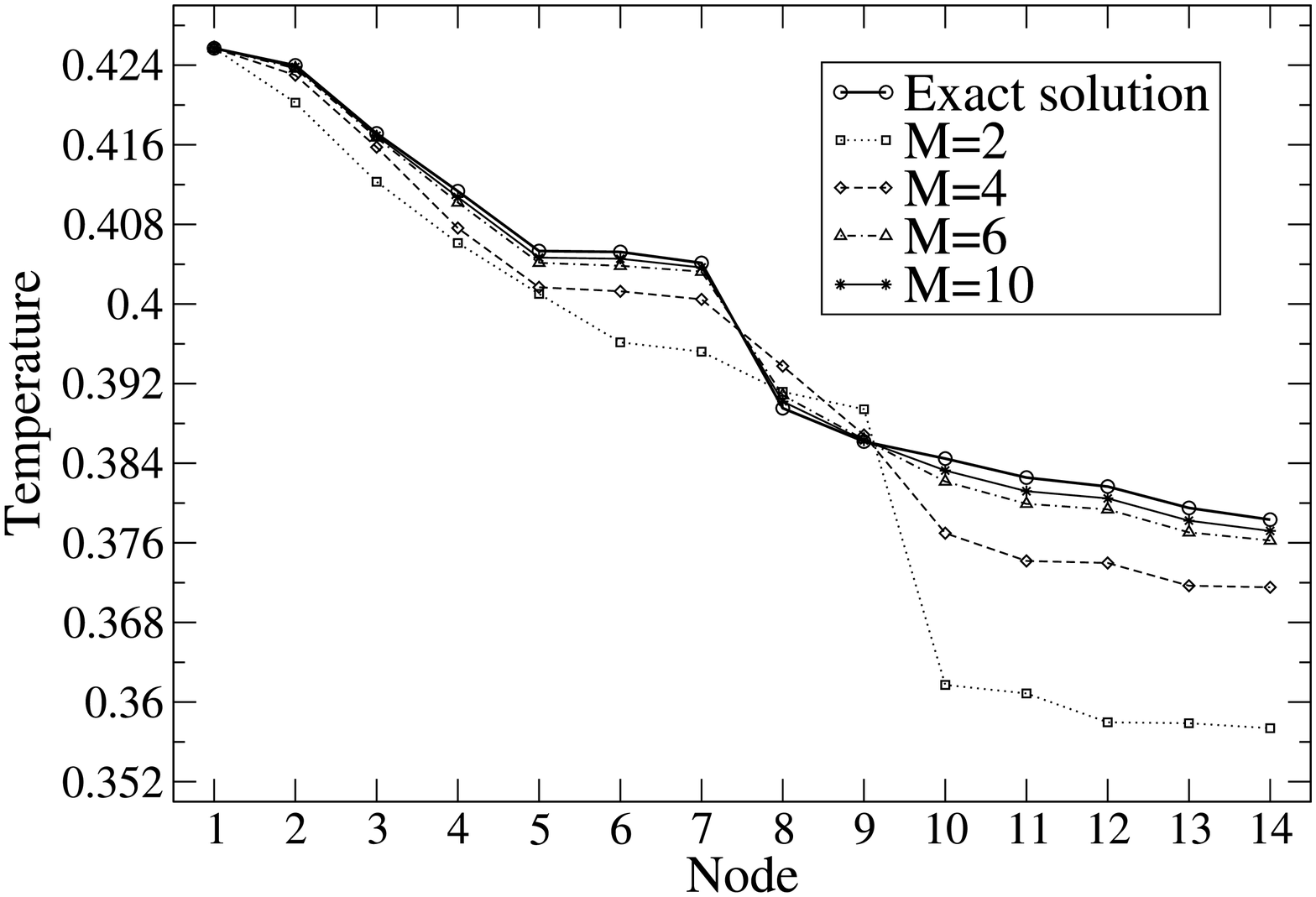} 
\vspace{-0.4in}
\caption{Comparison between the exact solution (bold line)
  eqs.(\ref{sol.3}-\ref{c.sol}) 
 and our approximation. For both cases we plot the ``hottest'' nodes.
For better visualization we shifted the profiles such that
the first node coincide in the graph.  We observe a good agreement
between the exact solution and the approximation for $M=10$ in our
artificial network. \vspace{-0.2in}
} \label{fig:comp} 
\end{center} 
\end{figure}

 The convergence of the approximate solution to the exact 
 solution (eqs.(\ref{sol.3}-\ref{c.sol})) was first tested 
  on an artificially generated random network of $100$ nodes.   
  Aside from the condition that the 
 nodes do not form disjoint clusters, a pair of nodes has 
 probability $p=0.1$ to be connected. One then randomly 
  selects a sink node and a source node that are 
 not directly linked.
 We expect to get very similar shape of the temperature-profile as 
 in the exact case. This is because for the 
 row-normalized matrix, the $| u^0 \ra$ vector being a column vector
 with $1$ in each entry may induce a small but
  {\it uniform} offset in the approximate solution.  
  In Fig.~\ref{fig:comp}, we plot the ``temperature-profile'' 
 of the $15$ hottest nodes from the exact
solution and the ``temperature-profile'' of the
same nodes using our approximation solution of various $M$.
 A good agreement between the exact solution and the approximate solution
 is reached at about $M=10$. 

To test the usability of our approach in real world, 
 we use the movielens database.  {\bf MovieLens} 
(movielens.umn.edu; grouplens.org)    
ratings are recorded on a five stars scale and contain additional
information, such as the time at which an evaluation was made.
The data set we downloaded contains $N=6040$ users $\times$ $M=3952$
movies. However, only a fraction $\xi_M=0.041$ of all possible votes
were actually expressed. To be able to perform the calculation in
 reasonable time, we decide to further reduce the data size in
 each dimension by roughly $50$\%. To preserve the 
 statistical properties of the original data, the pruning is
 done randomly without bias.  In
particular, we tried to maintain the probability distribution of the
number of votes per users, as well as the sparsity and the $N/M$
ratio. We want to stress that this is crucial when testing the
performance of predictive algorithms on real data in an objective
way. In fact, many  recommender systems can be found in the literature
 that rely on dense voting matrices~\cite{dense1,dense2}, at least in the
 traning data set. 
 Typically, users who have judged too few items are struck out, as well as
items that have received too few votes.  We did not comply to such
convention and made an effort to {\em keep the filtering level as low
as possible}, although this makes predictions much more difficult.

\begin{figure}[t] 
\begin{center} 
\vspace{-0.25in}
\includegraphics[width=0.5\textwidth]{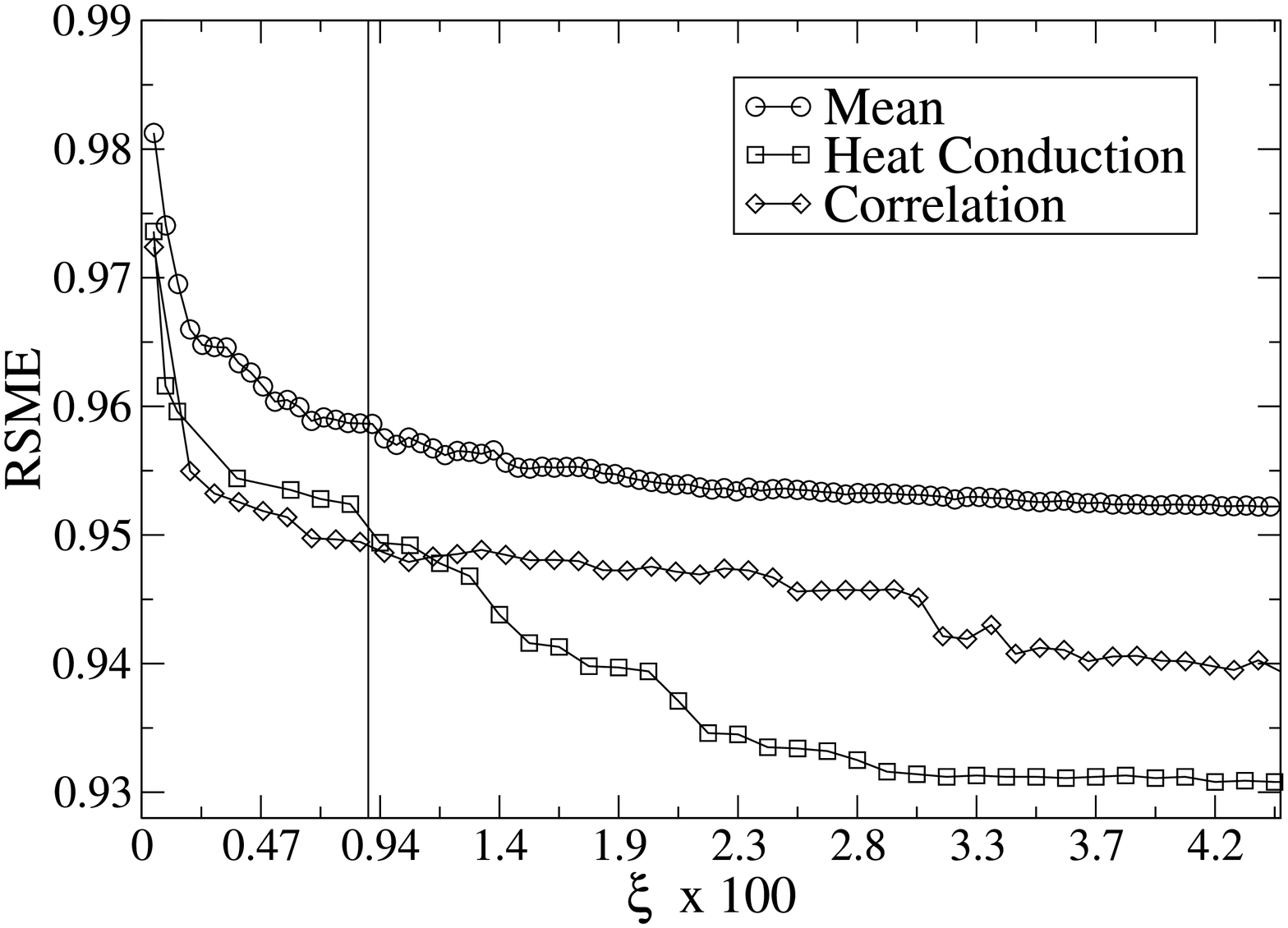} 
\vspace{-0.35in}
\caption{Prediction
performance on movielens database. The heat conduction model
outperforms the mean predictor and the Pearson correlation based
method as well. $\xi$ denotes the fraction of possible votes in the
matrix. The vertical line, corresponding approximately to 
 the giant cluster formation threshold in the movie -- movie network, 
  has vote density $\xi \approx 2N^{-1/2}M^{-1/6}$~\cite{Remark},
 where $N$ is the number of users, $M$ is the number of movies. 
\vspace{-0.25in}
} \label{fig:result} 
\end{center}
\end{figure}

Once filtered,  we cast the data set in a vote
matrix ${\mathbf V}$, with number of users $N=3020$ and
number of movies $M=1976$. In this reduced vote matrix, the 
 matrix element $V_{\alpha, i}$ represents the number of stars 
 assigned to movie $j$  by user $\alpha$ and is set to zero for
 unexpressed votes. The total 
filling fraction of ${\mathbf V}$ is $\xi_M = 0.0468$.  
{\em The votes in ${\mathbf V}$ are then sorted according
to their relative timestamps}. The last $n_{\rm test} = 10^4$ expressed
votes are collected to form our test set, while the rest of the 
expressed votes form our training set. We denote by ${\mathbf V}(t)$ 
 the vote matrix information up to time $t$. That is, in ${\mathbf V}(t)$
 all the unexpressed votes up to time $t$ are set to have zero star.

For the purpose of rating prediction, one will need a movie -- movie network. 
To accomplish this task, one may compute 
the correlation coefficient $C_{ij}(t)$ between movie $i$ and movie $j$ 
using the expressed votes up to a certain time $t$ in the training set. 
Specifically, we denote
 $\mu_i(t)\equiv {1\over N} \sum_{\alpha = 1}^N V_{\alpha, i}(t) $  
and $\sigma_i^2(t) \equiv {1\over N} \sum_{\alpha = 1}^N [V_{\alpha, i}(t)-\mu_i(t)]^2 $.
 The correlation coefficient reads
\be
C_{ij}(t) \equiv {\sum_\alpha [V_{\alpha,i}(t)-\mu_i(t)][V_{\alpha,j}(t)-\mu_j(t)]
 \over \sigma_i(t) \sigma_j(t)}\; .
\ee  
With a specified cutoff $C_{\rm cut}$, one obtains an adjacency matrix $A(t)$, with 
$A_{ij}(t) = \theta(C_{ij}(t)- C_{\rm cut}(t))$. 
The value of $C_{\rm cut}(t)$ is set 
so that the average degree per node $k(t)$ 
 for the movie -- movie
network has
 the same number of non-zero entries as  
 $[{\mathbf V}(t)]^T[{\mathbf V}(t)]$.

Keeping the test set data
fixed, 
we progressively fill the vote matrix the training set data over
time (using the relative time stamps), say up to time $t$. 
We then use $A(t)$ to construct the the propagator $D(t)$ 
based on the information accumulated up to $t$.  
For each viewer (user), 
the BC is simply given by the votes expressed by
the user up to time $t$. In the event that a user only has
one vote (or none) up to time $t$, the BC for that user
is given by randomly choosing one (or two) movie(s) and use the 
average rating(s) of the movie(s) up to that time as the boundary 
values~\cite{Remark_bc}.   
We then use our algorithm to make predictions on the entire test set. 
 
This test protocol is intended to reproduce real application tasks,
where one aims to predict future votes --which is, of course, much
harder than predicting randomly picked evaluations. It is somewhat
less realistic to fix the test set once and for all, but this has the
advantage to allow for more objective comparisons of the results.
Many different accuracy metrics have been proposed to assess
the quality of recommendations (see ref.~\cite{HeKo04}),
we choose the Root Square Mean Error:
\begin{equation}\label{mae} 
RSME=\sqrt{ 
\sum_{(\beta,j) \in \rm test}
 (V_{\beta,j}^{'} - V_{\beta,j})^{2}/n_{\rm test} }, 
\end{equation} 
where $V'_{\beta,j}$ represents the predicted vote from our algorithm, 
 $V_{\beta,j}$ represents the actual vote (rated by user $\beta$ 
 on movie $j$) in the test set, 
 and the sum runs over all expressed votes in the test set. 
In our experiments,
the RSME is calculated, at different sparsity values $\xi$, on a
unique test set.

Fig.~\ref{fig:result} summarizes the performance comparison  
of our model with the mean predictor
(the prediction is simply given by the objects mean value) and the 
widely used Pearson correlation based method \cite{ReIa94,HeKo00}.
Our model outperforms both after enough votes (of the order of
 $N^{1/2}M^{5/6}$) have been expressed. Since the dimensions of the vote
 matrix ${\mathbf V}$ is known in a real application, given the number 
 of expressed votes,
 it is relatively easy to see where one stands in terms of information content
 and whether our method will perform well using the given partial information.

In summary, we have devised a recommendation mechanism using analog to
 heat conduction. The innovation of our method is its capability to
 compute the Green's function needed just once to accommodate all possible
 BC. In terms of generalization, 
 it is apparent that our method can be applied to network with 
 weighted edges, with  $A_{ij}=w_{ij} \ge 0$. Whether such a generalization
 will improve the performance will be investigated in a separate 
 publication. Finally, we stress that 
 our study is not aimed to extract statistical properties out of networks
 through constructing model networks mimicking the real world 
 networks~\cite{New03,PaNew04}; nor are we pursuing  analysis of slowly
 decaying eigenmodes~\cite{Maslov_Sneppen} in the 
absence of boundary condtitions. Instead, our goal is to provide a framework
 that is capable of providing {\it individualized} information extraction from
 a real world network. 

YCZ and MB were partially supported by Swiss National Science 
Foundation grant 205120-113842. YCZ acknowledges hospitality at
 Management School, UESTC, China, where part of the work is done.
The research of YKY was supported by the Intramural Research Program of
the National Library of Medicine at the NIH.


\begin{thebibliography}{17}
\bibitem[{\citenamefont{Bollob\'as}(2001)}]{Remark_bc}
This is to avoid the artifact of null information retrieval: e.g.  
 assume only one boundary node with a specified temperature, all nodes
 will reach the same temperature upon thermal equilibrium.
\end{thebibliography}

\end{document}